\documentclass[twocolumn]{aastex631}

\submitjournal{ApJ}
\usepackage{cuted}
\usepackage{amsmath}
\usepackage{booktabs}
\usepackage{float}
\usepackage{longtable}
\usepackage{cuted}
\usepackage{placeins}
\usepackage{enumitem}
\setlist[enumerate]{itemsep=0mm}

\begin{document}
\title{The 230\,GHz Variability of Numerical Models of Sagittarius~A* \\ II. The Physical Origins of the Variability}

\author[0000-0003-2776-082X]{Ho-Sang Chan}
\email{hschanastrophy1997@gmail.com}
\altaffiliation{Croucher Scholar}
\affiliation{JILA, University of Colorado and National Institute of Standards and Technology, 440 UCB, Boulder, CO 80309-0440, USA}
\affiliation{Department of Astrophysical and Planetary Sciences, University of Colorado, 391 UCB, Boulder, CO 80309, USA}

\author[0000-0001-6337-6126]{Chi-kwan Chan}
\affiliation{Steward Observatory and Department of Astronomy, University of Arizona, 933 N. Cherry Avenue, Tucson, AZ 85721, USA}
\affiliation{Data Science Institute, University of Arizona, 1230 N. Cherry Avenue, Tucson, AZ 85721, USA}
\affiliation{Program in Applied Mathematics, University of Arizona, 617 North Santa Rita, Tucson, AZ 85721, USA}

%\author[0000-0002-0393-7734]{Ben S. Prather}
%\affiliation{Los Alamos National Lab, Los Alamos, NM, 87545, USA}

%\author[0000-0001-6952-2147]{George N. Wong}
%\affiliation{School of Natural Sciences, Institute for Advanced Study, 1 Einstein Drive, Princeton, NJ 08540, USA}
%\affiliation{Princeton Gravity Initiative, Princeton University, Princeton, NJ 08544, USA}

%\author[0000-0001-7451-8935]{Charles Gammie}
%\affiliation{Department of Physics, University of Illinois at Urbana-Champaign, 1110 West Green Street, Urbana, IL 61801, USA}
%\affiliation{Department of Astronomy, University of Illinois at Urbana-Champaign, 1002 West Green Street, Urbana, IL 61801, USA}
%\affiliation{NCSA, University of Illinois at Urbana-Champaign, 1205 W. Clark St., Urbana, IL 61801, USA}
%\affiliation{Illinois Center for the Advanced Study of the Universe, University of Illinois at Urbana-Champaign, 1110 West Green St., Urbana, IL 61801, USA}

%%%%%%%%%%%%%%%%%%%%%%%%%%%%%%%%%%%%%%%%%%%%%%%%%%%%%%%%%%%%%%%%%%%%%%%%%%%%%%%%%%%%%%%%%%%%%%%%%%%%%%%%%%%%%%%%%%%%%

\begin{abstract}

%In \citet{2024ApJ...964...17C}, we explored how the ion-electron temperature ratio affects certain numerical models of Sagittarius~A* (Sgr~A*). Specifically, we studied these effects in a magnetic-arrested disk (MAD), focusing on the $3$-hour variability at $230$\,GHz ($M_{\Delta T}$). 

We continue our previous work, \citet{2024ApJ...964...17C}, to investigate how variations in the electron temperature prescription parameter, $R_{\rm Low}$, influence the $3$-hour variability at $230$\,GHz, $M_{\Delta T}$, in magnetic-arrested disk (MAD) models of Sagittarius~A* (Sgr~A*), through analyzing a series of general-relativistic magnetohydrodynamics and raytracing simulations. For models with a black hole spin $a > 0$, we discovered that increasing $R_{\rm Low}$ renders the photon ring more optically thick, obscuring the varying accretion flows that contribute to the variability. However, as $R_{\rm Low}$ increases further, MAD flux eruptions become more pronounced, compensating for the decrease in $M_{\Delta T}$. For models with a spin $a < 0$, although a higher $R_{\rm Low}$ also increases the optical thickness of the fluid, voids within the optically thick gas fail to cover the entire photon ring. Similarly, flux eruptions become more prominent as $R_{\rm Low}$ increases further, contributing to the observed rise in $M_{\Delta T}$ relative to $R_{\rm Low}$. For black holes with a spin $a = 0$, although the effect of increasing optical depth is still present, their $230$\,GHz light curves, and hence $M_{\Delta T}$, are insensitive to changes in $R_{\rm Low}$.  Furthermore, we found that the variability of the $230$\,GHz light curves at $R_{\rm Low} = 1$ might correlate with fluctuations in the internal energy of the gas near the black hole, and we listed potential causes and solutions to the over-variability problem. Our findings highlight potential approaches for refining $M_{\Delta T}$ to better align with observations when modeling Sgr~A*.

\end{abstract}
 

%%%%%%%%%%%%%%%%%%%%%%%%%%%%%%%%%%%%%%%%%%%%%%%%%%%%%%%%%%%%%%%%%%%%%%%%%%%%%%%%%%%%%%%%%%%%%%%%%%%%%%%%%%%%%%%%%%%%%

\section{Introduction} \label{sec:intro}

The galactic-center supermassive black hole Sagittarius A* (Sgr~A*) is estimated to accrete matter at a rate of $\sim 10^{-5}$ relative to its Eddington limit, $\dot{M}_{\rm Edd}$ \citep{yuan2002jet}. Under such conditions, the accretion disk around Sgr~A* can be modeled as an advection-dominated, radiatively inefficient accretion flow \citep[RIAF,][]{quataert2003radiatively, begelman2012radiatively}, where the accreting plasma is optically thin and geometrically thick. While ions are more effectively energized via viscous dissipation \citep{yuan2014hot}, electrons are the main channels for radiative cooling \citep{begelman2014accreting}. The poor Coulomb coupling between ions and electrons prevents ions from transferring their energy to electrons and transporting it outward through radiation. Even if viscous dissipation can heat electrons, they cool much more rapidly than ions \citep{begelman2014accreting}. As a result, most of the energy is advected into the black hole. This collisionless nature of the plasma naturally leads to a two-temperature state between the electrons and ions, where $T_{e}$ (electron temperature) $\neq$ $T_{i}$ (ion temperature). %Additionally, electrons could be accelerated to a non-thermal \citep{uzdensky2022relativistic} distribution presumably through magnetic reconnections \citep{werner2018non} and turbulence \citep{zhdankin2019electron}, making it even less likely to have $T_{e} \approx T_{i}$. The bulk ion plasma is hot with respect to its viral temperature \citep{narayan1994advection, blandford1999fate}.

An accurate description of the electron temperature is essential for reproducing and matching the electromagnetic observables of Sgr~A*. However, such an accurate treatment requires either two-temperature simulations that separately evolve the electron entropy \citep{ressler2015electron, ressler2017disc, chael2018role, chael2019two, dexter2020parameter, scepi2022sgr, grigorian2024relationship}, or first-principles, particle-in-cell plasma simulations in the general-relativistic limit \citep{2016ApJ...826...77B, galishnikova2023collisionless}, which are both computationally expensive. An ad-hoc approach is to assign $T_{e}$ based on the local $T_{i}$ \citep{dexter2010submillimeter}. For instance, \citet{2009ApJ...703L.142D} presented a synchrotron emission model of Sgr~A* images in which $T_{e} = T_{i}$ everywhere; \citet{2009ApJ...706..497M} considered $T_{i}$ to be a fraction $R$ of $T_{e}$ and compared model spectral energy distributions with observations; \citet{2013A&A...559L...3M} allows $R$ to vary across the domain, so that $R$ in the jet is different from that in the disk; \citet{chan2015power} proposed a prescription function in which $R$ is a step-function of the plasma $\beta$ ($\beta = 2P/b^{2}$, where $P$ is the gas pressure and $b^{2}$ is the square of the magnetic field strength) at some threshold $\beta_{\rm Crit}$ (usually set as $1$); \citet{moscibrodzka2016general} proposed the prescription function:
\begin{equation} \label{eqn:fraction}
    R = \frac{T_{i}}{T_{e}} = \frac{R_{\rm High}b^{2} + R_{\rm Low}}{b^{2} + 1},
\end{equation}
where $R_{\rm High}$ ($R_{\rm Low}$) represents the ion-electron temperature ratios in the weakly (strongly) magnetized regime. This function smooths the sharp transition of the step-function prescription by \citet{chan2015power}, and the power-law dependencies on $b$ guarantee that radiations from the strong and weak ion-electron coupling regions are easily distinguishable. Although still ad-hoc, both prescriptions are simple and are physically motivated by the results of particle-in-cell simulations showing that collisionless plasma preferentially heats the ions for $\beta > 1$ \citep{akiyama2019first}. While this prescription function lacks detailed microphysics of the viscous heating ratio between the ions and the electrons, one can easily tune the free parameters to reproduce a wide range of spectra and images that can be compared with observations and constrain the parameter spaces. Note that the electron temperature is given as:
\begin{equation}
    T_{e} = \frac{2m_{p}\epsilon}{3k(2+R)}
\end{equation}
where $\epsilon$ is the specific internal energy of the bulk flow, and we assume that the plasma is an electron-proton plasma such that their internal energies add up to the bulk internal energy. Note that we also assume the electron (proton) is relativistic (non-relativistic) with an adiabatic index of $4/3$ ($5/3$).

This simple description was used in \citet{2022ApJ...930L..16E} to compare the $230$\,GHz flux between observations and those generated through general-relativistic magnetohydrodynamics (GRMHD) simulations and general-relativistic ray-tracing (GRRT) post-processing. None of their models simultaneously pass all of the observational constraints, including the $230$\,GHz flux variability, which is defined as the ratio $M_{\Delta T} = \sigma_{\Delta T}/\mu_{\Delta T}$, where $\sigma_{\Delta T}$ is the standard deviation and $\mu_{\Delta T}$ is the mean of the $230$\,GHz flux over a time $\Delta T$. For the EHT Collaboration study, $\Delta T = 530\,GMc^{-3} = 3\,\mathrm{hours}$. The reason why some theoretical models are more variable than observations remains uncertain. In \citet{2024ApJ...964...17C}, we try to address this problem by varying another free parameter $R_{\rm Low}$ in Equation \ref{eqn:fraction}, where it is usually assumed to be $1$. Here, we summarize our findings from \citet{2024ApJ...964...17C}:
\begin{itemize}
    \item Increasing $R_{\rm Low}$ can lead to a reduction in $M_{\Delta T}$, but it is model-dependent 
    
    \item Most models with $a > 0$ show a first decrease then increase trend in $M_{\Delta T}$ when one increases $R_{\rm Low}$
    
    \item Most models with $a < 0$ shows an increasing tendency in $M_{\Delta T}$ when one increases $R_{\rm Low}$

    \item Most models with $a = 0$ seems to be insensitive in $M_{\Delta T}$ when one changes $R_{\rm Low}$
    
    \item The major contribution to the high $M_{\Delta T}$ at $R_{\rm Low} = 1$ is from the photon ring
\end{itemize}
and we refer readers to Figure \ref{fig:paper1recap} for a recap on the parameter dependence of $M_{\Delta T}$ against $R_{\rm Low}$. We also remind readers here that the photon ring of a black hole image consists of an infinite series of sub-rings that characterize the number of orbits that a photon completes around the black hole before reaching the observer. Although models with $R_{\rm Low} \neq 1$ have been briefly explored in the contexts of M87 and Sgr~A* \citep{2021ApJ...910L..13E, 2022ApJ...930L..16E, 2023ApJ...957L..20E, 2025A&A...693A.265E}, a systematic parameter search and analysis on a wide range of $R_{\rm Low}$ has only been conducted in our previous work \citet{2024ApJ...964...17C}, focusing on Sgr~A*. However, we have not addressed the parameter dependence of $M_{\Delta T}$ on $R_{\rm Low}$. In particular:
\begin{itemize}
    \item Why does increasing $R_{\rm Low}$ lead to a reduction in $M_{\Delta T}$ in some but not all cases?

    \item What is the physical origin of the high $M_{\Delta T}$ at $R_{\rm Low} = 1$?
\end{itemize}
and we aim to examine these questions through an in-depth analysis of the images and simulation snapshots. This paper is structured as follows: We review the procedure of constructing the GRMHD simulation libraries and the GRRT parameter surveys in Section \ref{sec:method}, highlighting the parameter spaces we considered in this study. We also describe our process of selecting representative models to address the above questions. In Section \ref{sec:results}, we analyze the parameter dependence of $M_{\Delta T}$ on $R_{\rm Low}$ by examining the images of different $R_{\rm Low}$ and the corresponding GRMHD snapshots. As we will see, the reduction of $M_{\Delta T}$ is an optical depth effect. We discuss the implications of our study in Section \ref{sec:discuss}, and we conclude our studies in Section \ref{sec:conclu}.

%%%%%%%%%%%%%%%%%%%%%%%%%%%%%%%%%%%%%%%%%%%%%%%%%%%%%%%%%%%%%%%%%%%%%%%%%%%%%%%%%%%%%%%%%%%%%%%%%%%%%%%%%%%%%%%%%%%%%

\section{Methodology} \label{sec:method}

%%%%%%%%%%%%%%%%%%%%%%%%%%%%%%%%%%%%%%%%%%%%%%%%%%%%%%%%%%%%%%%%%%%%%%%%%%%%%%%%%%%%%%%%%%%%%%%%%%%%%%%%%%%%%%%%%%%%%

\subsection{GRMHD Simulations} \label{sec:grmhd}

We briefly review the procedure of constructing the GRMHD simulation libraries and the parameter surveys for GRRT images of varying $R_{\rm Low}$. The simulation library (named model v3, see, for instance, Table 1 of \citet{2022ApJ...930L..16E}) is obtained using the open-source code kharma \citep{prather2021iharm3d}. kharma solves the ideal GRMHD equations (with $c = 1$):
\begin{equation}
\begin{aligned}
    \partial_{t}(\sqrt{-g}\rho u^{t}) &= -\partial_{i}(\sqrt{-g}\rho u^{i}), \\
    \partial_{t}(\sqrt{-g}T^{t}_{\;\;\nu}) &= -\partial_{i}(\sqrt{-g} T^{i}_{\;\;\nu}) + \sqrt{-g}T^{\kappa}_{\;\;\lambda}\Gamma^{\lambda}_{\;\;\nu\kappa}, \\
    \partial_{t}(\sqrt{-g}B^{i}) &= -\partial_{j}[\sqrt{-g}(b^{j}u^{i} - b^{i}u^{j})], \\
    \partial_{i}(\sqrt{-g}B^{i}) &= 0.
\end{aligned}
\end{equation}
and here, $g_{\mu\nu}$ is the Kerr metric in the funky modified Kerr-Schild coordinates \citep{wong2022patoka}, $g$ is the determinant of the metric, $\rho$ is the mass density, $\Gamma^\lambda_{\nu\kappa}$ is the Christoffel symbol, $B^{i}$ is the magnetic field in the coordinate frame, $u^{\mu}$ is the $4$-velocity, $b^{\nu}$ is the magnetic $4$-vector, and $T^{\mu\nu}$ is the stress-energy tensor \citep{dhang2023magnetic} defined by:
\begin{equation}
    T^{\mu\nu} = (\rho h + b^{2})u^{\mu}u^{\nu} + (P + \frac{b^{2}}{2})g^{\mu\nu} - b^{\mu}b^{\nu}.
\end{equation}
so that $h = 1 + \gamma/(\gamma - 1)P/\rho$ is the specific enthalpy. The $\nabla \cdot \vec{B} = 0$ constraint is maintained by the flux-constrained transport scheme \citep{toth2000b}. kharma uses the 5th order Weighted-Essential-Non-Oscillatory scheme \citep{jiang1996efficient} to reconstruct primitive variables to the cell boundaries and the Lax-Friedrichs solver to construct the Riemann fluxes. kharma evolves the GRMHD equations using the 2nd order Strong Stability Preserving Runge-Kutta Method \citep{gottlieb2011strong} with a Courant–Friedrichs–Lewy number of $0.7$.

In this work and \citet{2024ApJ...964...17C}, we consider only the kharma simulations of the Magnetic Arrested Disc (MAD) state since the MAD models are favored in \citet{2022ApJ...930L..16E}. Additionally, recent near-horizon circular polarization measurements and model comparisons favor Sgr~A* to have dynamically important magnetic fields \citep{2024ApJ...964L..26E}. The accreting plasma assumed an ideal gas equation of state with an ion adiabatic index of $4/3$, and the initial conditions were based on the Fishbone-Moncrief torus \citep{1976ApJ...207..962F}. The torus has an inner radius at $20$ and a pressure maximum at $41$, all in units of $GMc^{-2}$. To initialize the MAD simulation, the initial magnetic field within the torus is set via the vector potential \citep{wong2022patoka}:
\begin{equation} \label{eqn:mad}
    A_{\phi} \propto \text{max} \left[\frac{\rho}{\rho_{\rm max}}\left(\frac{r}{r_{0}}\text{sin} \theta\right)^{3}e^{-r/400} - 0.2, 0 \right],
\end{equation} 
where $\rho_{\rm max}$ is the maximum density, $r$ is the radial coordinate, $r_{0}$ is the inner radial boundary of the computational domain, and $\theta$ is the polar angle (only in Equation \ref{eqn:mad}). The radial outer boundary is at $1,000$\,$GMc^{-2}$ with outflowing boundary conditions. The inner boundary is placed within a few blocks of the black hole horizon $r_{\rm BH} = 1 + \sqrt{1 - a^{2}}$ with $a$ being the spin of the black hole. Matter is allowed to flow into the horizon but not vice versa. Axis-symmetric boundary conditions are employed along the pole, while periodic boundary conditions are assumed along the azimuthal direction. The simulation resolution is $288 \times 128 \times 128$, and the simulation duration is 30,000\,$GMc^{-3}$, sufficient to ensure inflow equilibrium up to the radius of our interest.

%%%%%%%%%%%%%%%%%%%%%%%%%%%%%%%%%%%%%%%%%%%%%%%%%%%%%%%%%%%%%%%%%%%%%%%%%%%%%%%%%%%%%%%%%%%%%%%%%%%%%%%%%%%%%%%%%%%%%

\begin{figure*}[htb!]
    \centering
    \includegraphics[width=1.0\linewidth]{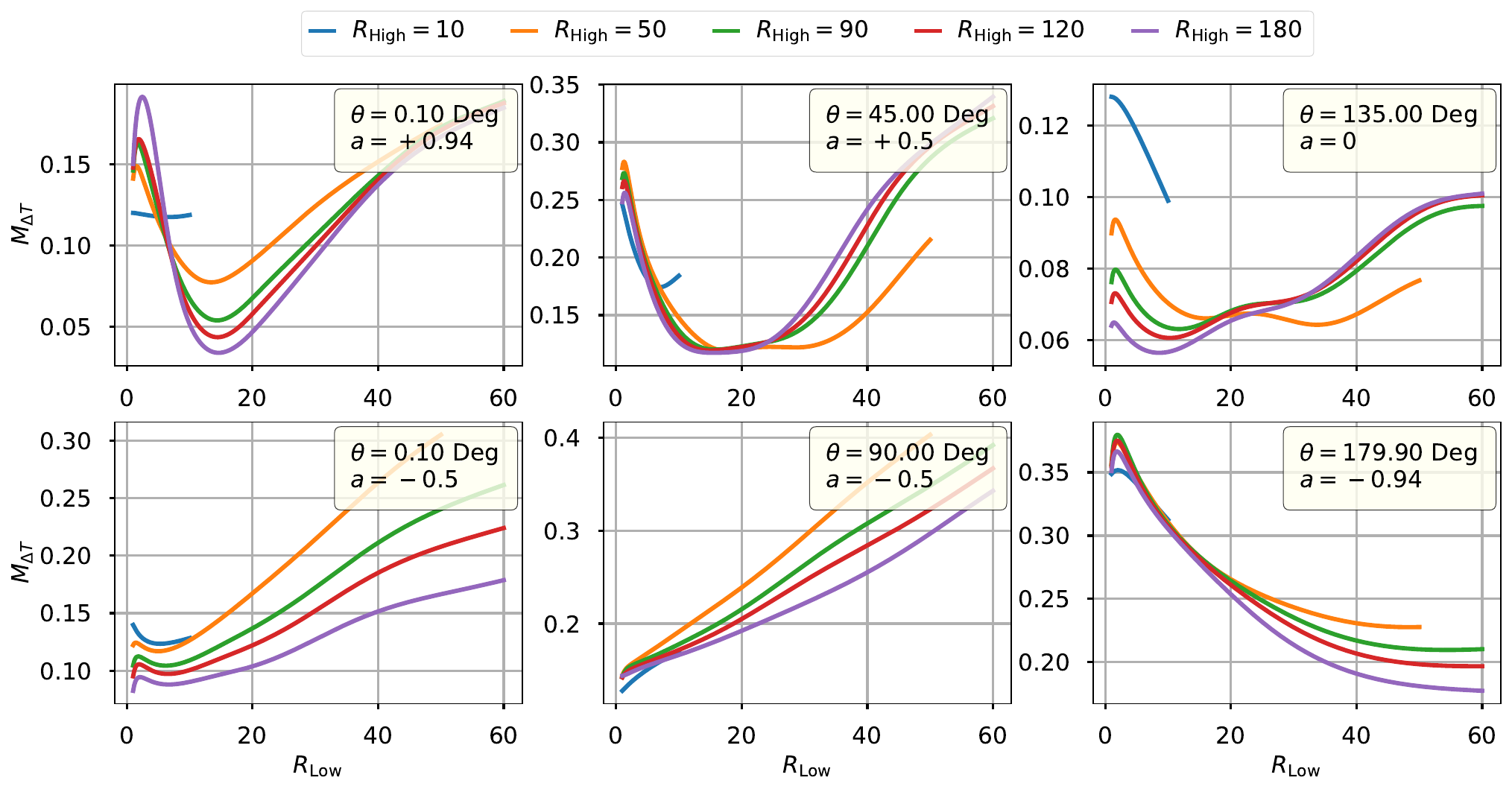}
    \caption{Recapping the parameter dependence of $M_{\Delta T}$ against $R_{\rm Low}$ from \citet{2024ApJ...964...17C}. We label the model $\theta$ and $a$ in the upper right-hand corner of each subplot. \label{fig:paper1recap}}
\end{figure*}

\begin{figure*}[htb!]
    \centering
    \includegraphics[width=1.0\linewidth]{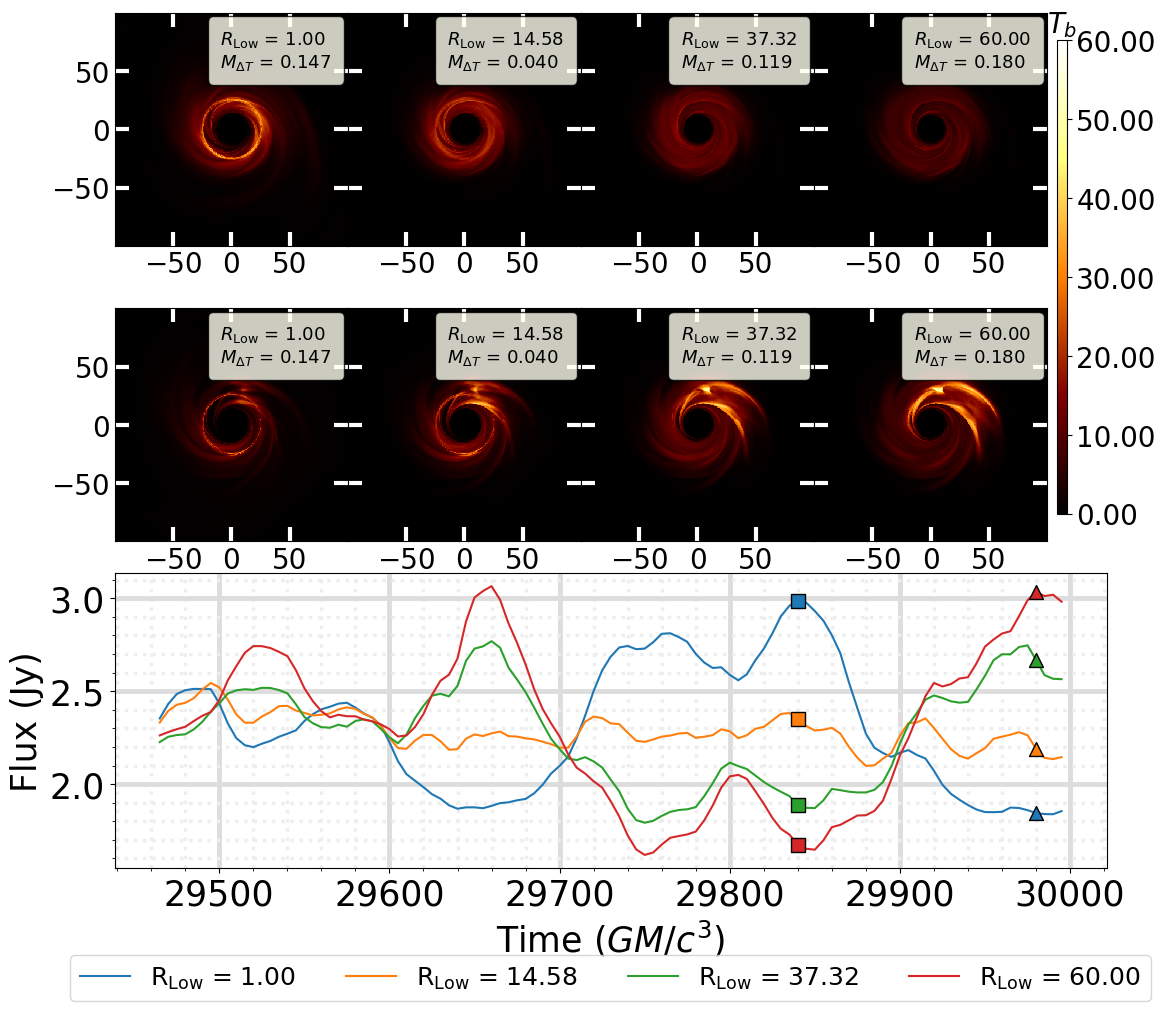}
    \caption{Lowest panel: the $230$\,GHz flux light curves for different $R_{\rm Low}$. Here, the black hole spin is $a = +0.94$, $\theta = 0.1$ Deg, and $R_{\rm High} = 180$. Upper-most panel: Black hole images with $R_{\rm Low}$ correspond to those shown in the lowest panel. In each subplot, we show the $R_{\rm Low}$ and $M_{\Delta T}$ in the upper right corner. The images are taken at the instants marked as squares on top of the light curve. The middle panel is the same as the upper-most panel, but the images are taken at the instants marked as triangles. Images are shown in units of brightness temperature $T_{b}$ (in $10^{9}$ K), and all images share the same color scale. Increasing $R_{\rm Low}$ blocks the photon ring from illuminating, thus reducing the variability. However, increasing $R_{\rm Low}$ further makes MAD flux eruptions more visible, hence increasing the variability again. \label{fig:a+0.94}}
\end{figure*}

\begin{figure}[htb!]
    \centering
    \includegraphics[width=1.0\linewidth]{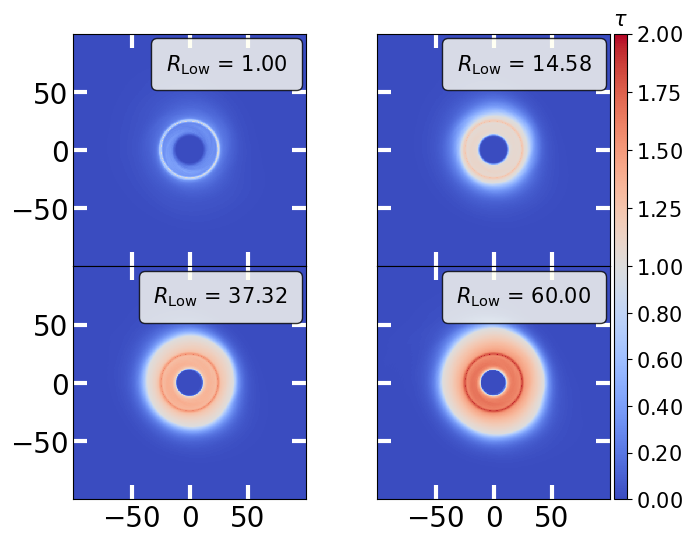}
    \caption{The time-averaged optical depth of the black hole with $a = +0.94$, $\theta = 0.1$ Deg, and $R_{\rm High} = 180$, but with varying $R_{\rm Low}$. Here, the optical depth between $0 - 1$ is in physical units, but those from $1 - 2$ are mapped via a power-law exponent $\gamma_{\rm map} = \text{log}_{10}(2)/\text{log}_{10}[\rm{MAX}(\tau)]$, so that the maximum $\tau$ across these four figures corresponds to $2$ in the color bar. The photon ring becomes optically thick as $R_{\rm Low}$ increases. The reduced variability is thus an optical depth effect. \label{fig:odepth}}
\end{figure}

\begin{figure}[htb!]
    \centering
    \includegraphics[width=1.0\linewidth]{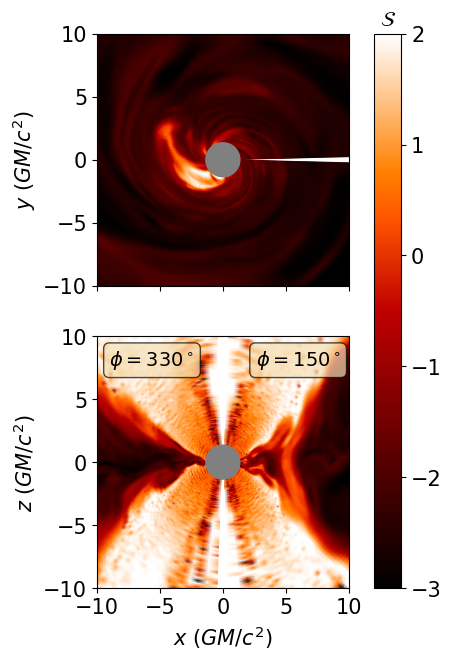}
    \caption{The entropy $\mathcal{S}$ (in the log$_{10}$ scale) contours along the $x-y$ ($x-z$) slice are shown in the upper (lower) panel, demonstrating the small-scale, MAD flux eruptions. This is for the black hole with $a = +0.94$ and is taken at the same instant as the triangular marker in Figure \ref{fig:a+0.94}. In the lower panel, we show the $\phi$ coordinate (in Deg) at the upper left- and right-hand corners, respectively. The MAD flux eruption in the image domain has the same morphology as that in the simulation domain, and it exhibits a tube-like structure. \label{fig:eruption}}
\end{figure}

\begin{figure*}[htb!]
    \centering
    \includegraphics[width=1.0\linewidth]{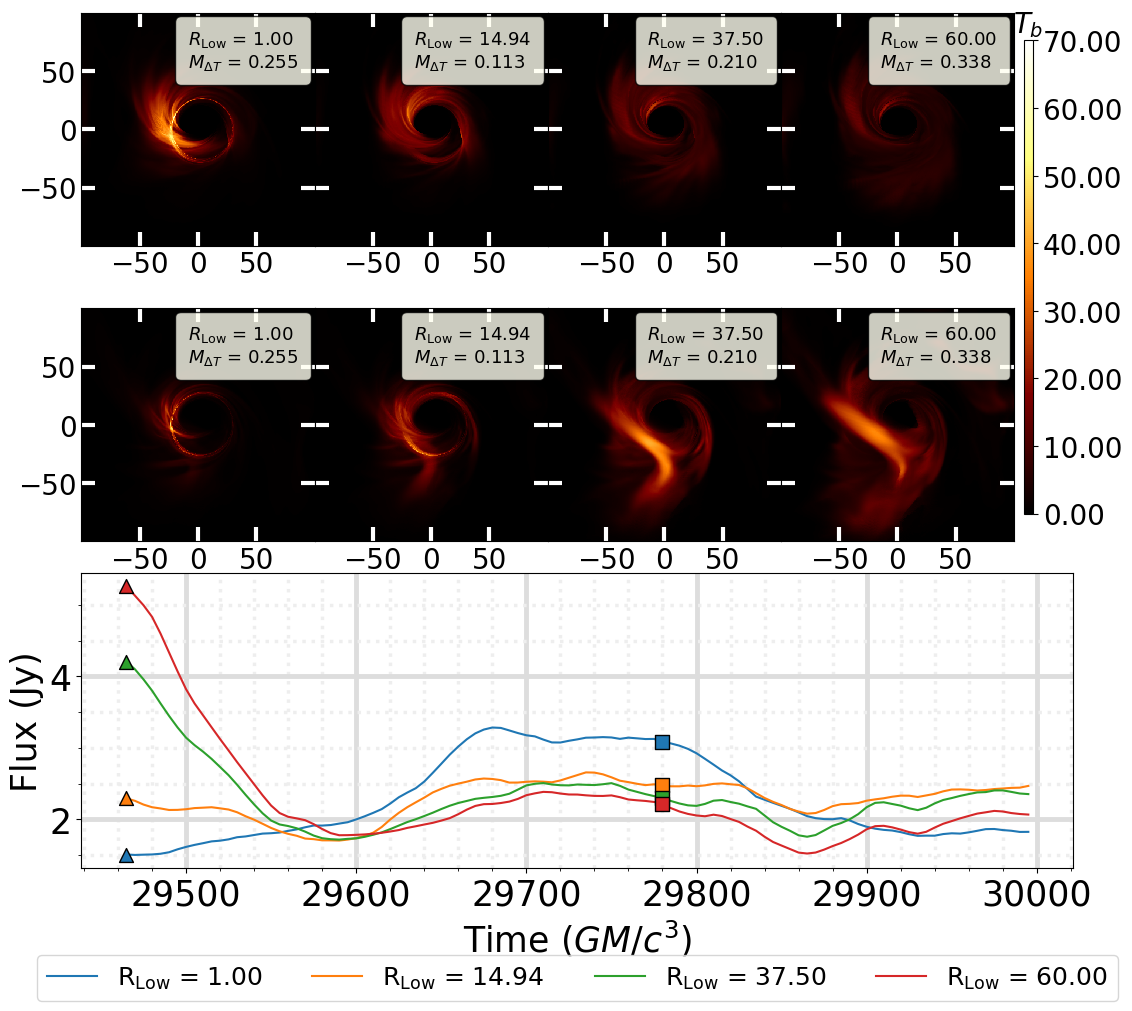}
    \caption{Same as Figure \ref{fig:a+0.94}, but for the black hole with $a = +0.5$, $\theta = 45$ Deg, and $R_{\rm High} = 120$. The enhanced variability at the beginning of this time chunk is manifested as a large-scale flux tube generated by the MAD flux eruption and becomes more visible as $R_{\rm Low}$ increases. \label{fig:a+0.5}}
\end{figure*}

\begin{figure}[htb!]
    \centering
    \includegraphics[width=1.0\linewidth]{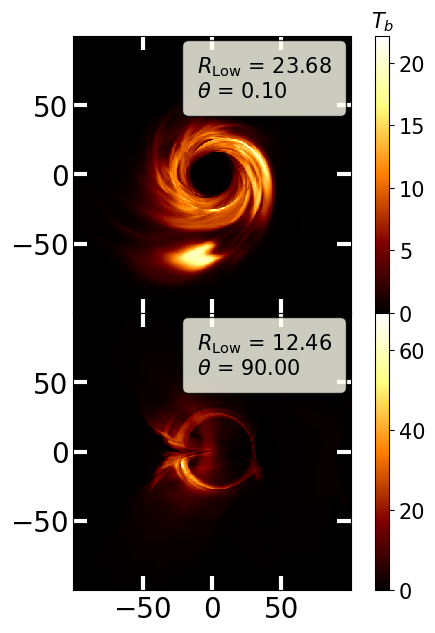}
    \caption{The magnetic flux tube for the black hole with $a = + 0.5$ and $R_{\rm High} = 120$ manifested in the $230$\,GHz image domain but captured from different viewing angles. Both images are shown in units of brightness temperature $T_{b}$ (in $10^{9}$ K). We show the $R_{\rm Low}$ and $\theta$ in the upper right corner of each subplot. This is taken at the instant marked as triangles in Figure \ref{fig:a-0.5}. \label{fig:fluxtube}}
\end{figure}

\begin{figure}[htb!]
    \centering
    \includegraphics[width=1.0\linewidth]{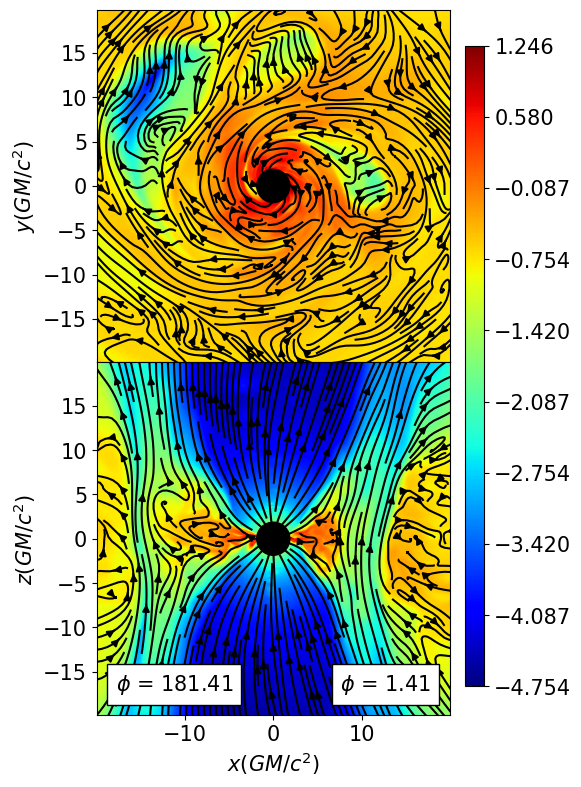}
    \caption{The magnetic flux tube shown in Figures \ref{fig:a+0.5} and \ref{fig:fluxtube} as manifested in the simulation domain. We show the $x-y$ ($x-z$) slice of the density contour (in the log$_{10}$ scale) with magnetic field lines in the upper (lower) panel. In the lower panel, we show the $\phi$ coordinate (in Deg) at the lower left- and right-hand corners, respectively. This is taken at the same instant as Figure \ref{fig:fluxtube}. \label{fig:fluxtube-vecfield}}
\end{figure}

\begin{figure*}[htb!]
    \centering
    \includegraphics[width=1.0\linewidth]{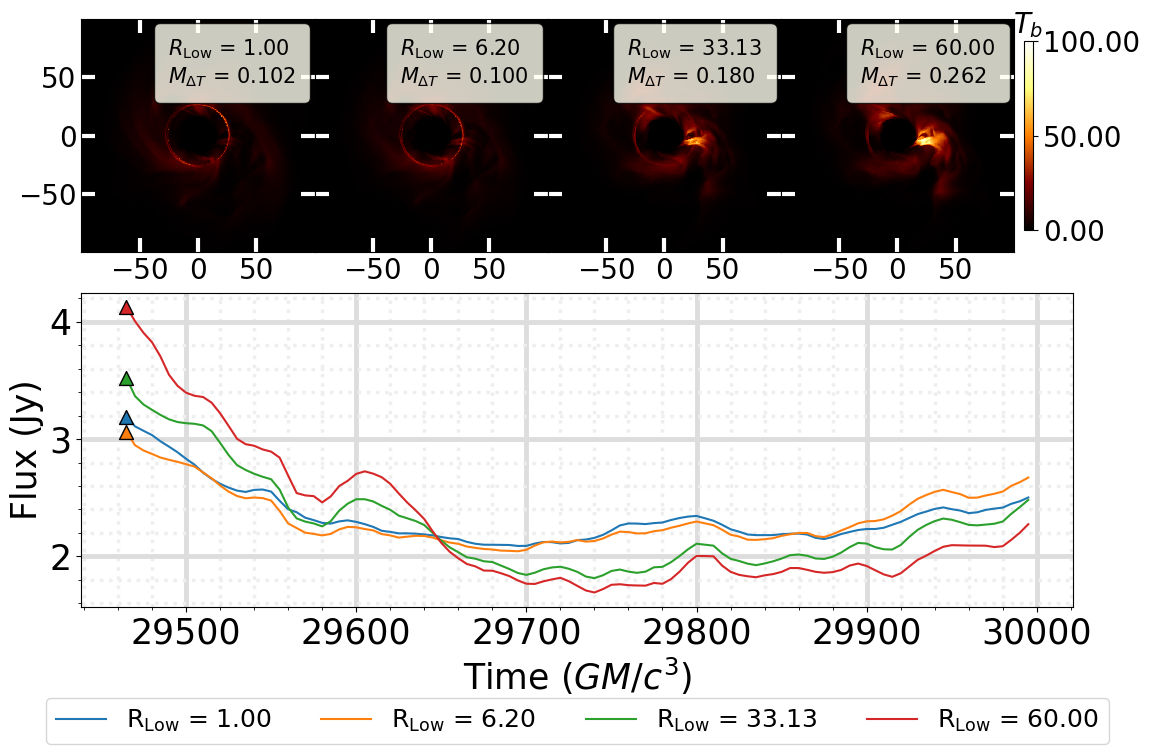}
    \caption{Same as Figure \ref{fig:a+0.94}, but for the black hole with $a = -0.5$, $\theta = 0.1$ Deg, and $R_{\rm High} = 90$. We show only the instant marked as triangles on the light curves. As $R_{\rm Low}$ increases, the gas becomes optically thick but contains voids that cannot entirely block the photon ring from illuminating. \label{fig:a-0.5}}
\end{figure*}

\begin{figure*}[htb!]
    \centering
    \includegraphics[width=1.0\linewidth]{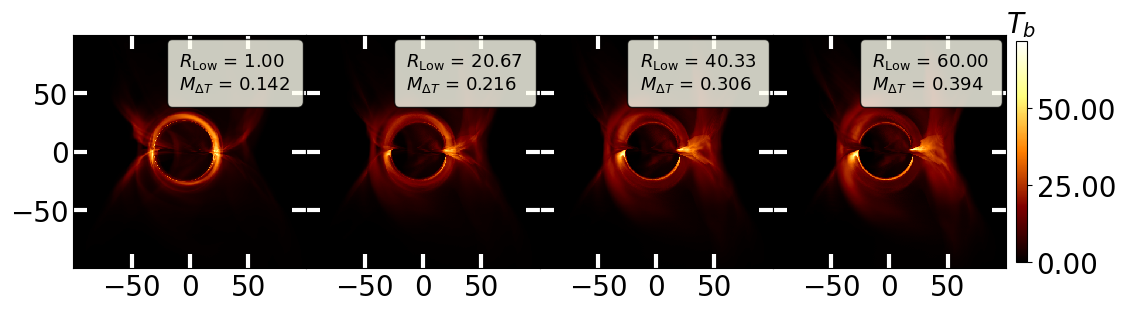}
    \caption{Same as Figure \ref{fig:a-0.5}, but with $\theta = 90$ Deg, and we omit the light curves for simplicity. The plasma is still unable to block the photon ring if the black hole is viewed from another angle. \label{fig:a-0.5-angle}}
\end{figure*}

\begin{figure}[htb!]
    \centering
    \includegraphics[width=1.0\linewidth]{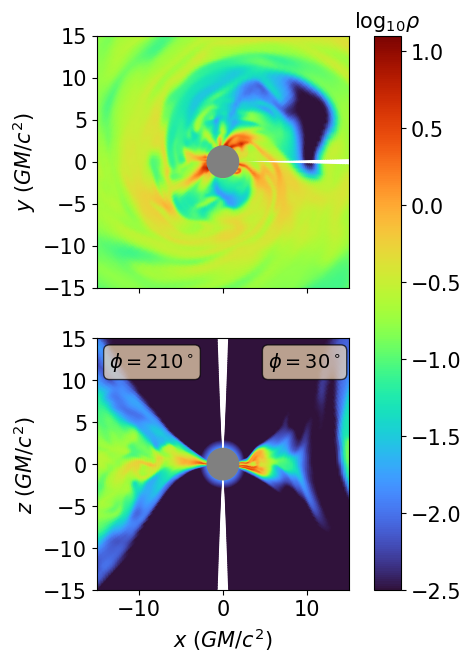}
    \caption{Same as Figure \ref{fig:fluxtube-vecfield}, but for the black hole with $a = -0.5$. This is to demonstrate the gas voids and flux eruptions as manifested in the simulation domain. We omitted vector streamlines for simplicity. This is taken at the same instant as Figure \ref{fig:a-0.5-angle}. \label{fig:void}}
\end{figure}

\begin{figure}[htb!]
    \centering
    \includegraphics[width=1.0\linewidth]{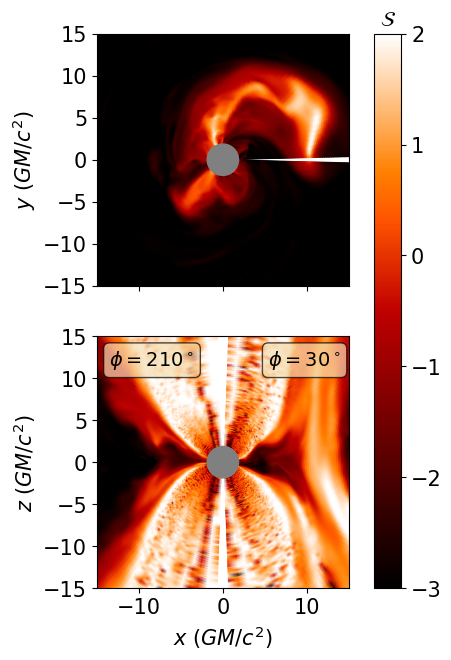}
    \caption{Same as Figure \ref{fig:eruption}, but for the black hole with $a = -0.5$. This is taken at the same instant as Figure \ref{fig:a-0.5-angle}, showing the MAD flux eruption morphology being consistent with the energetic transient found in the image domain. \label{fig:eruption-a-0.5}}
\end{figure}

\begin{figure*}[htb!]
    \centering
    \includegraphics[width=1.0\linewidth]{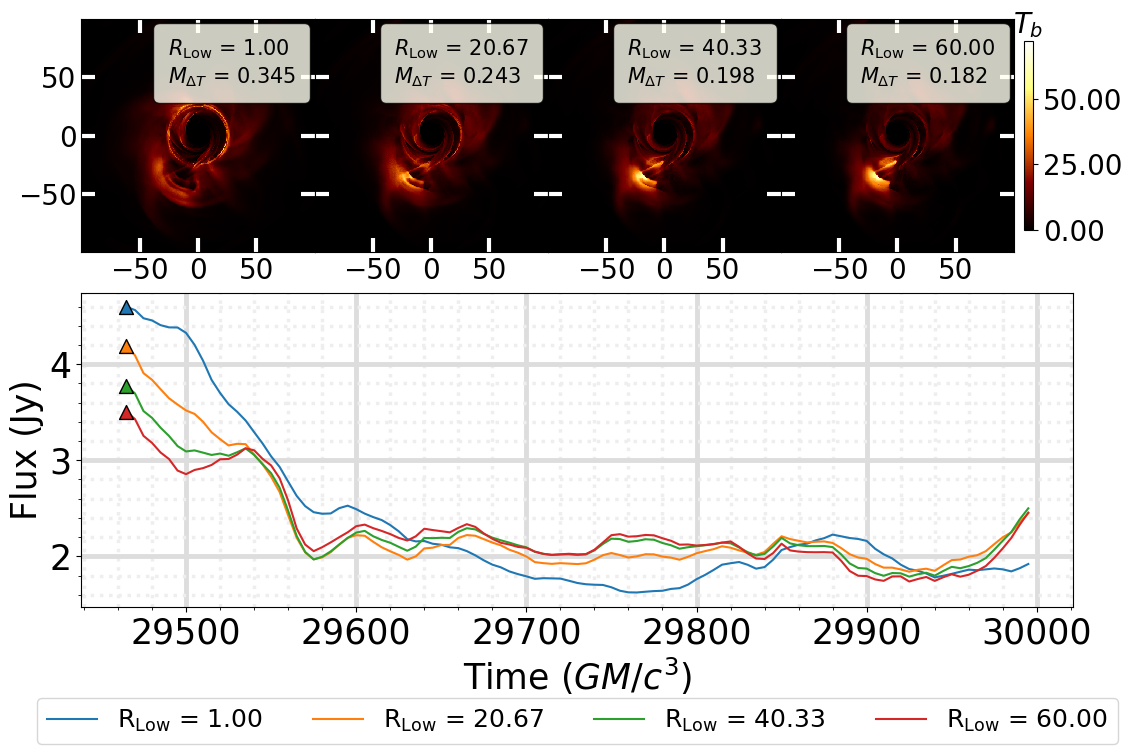}
    \caption{Same as Figure \ref{fig:a-0.5}, but for the black hole with $a = -0.94$, $\theta = 179.9$ Deg, and $R_{\rm High} = 180$. The increasing optical depth effect still applies, and the plasma contains voids, but the variability \textit{decreases} with $R_{\rm Low}$. We argue this should be a special case. \label{fig:a-0.94}}
\end{figure*}

\begin{figure*}[htb!]
    \centering
    \includegraphics[width=1.0\linewidth]{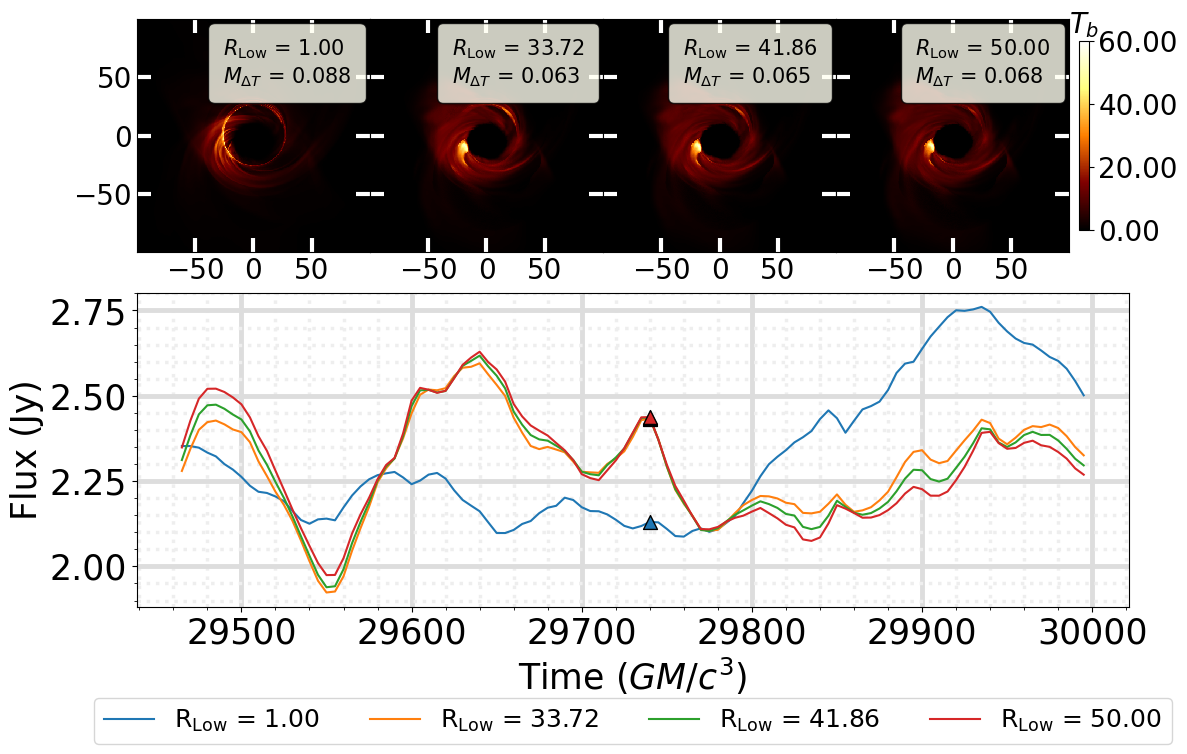}
    \caption{Same as Figure \ref{fig:a-0.5}, but for the black hole with $a = 0$, $\theta = 135$ Deg, and $R_{\rm High} = 50$. Here, we select the instant of the snapshot to show that the optical depth effect still applies, while increasing $R_{\rm Low}$ makes MAD flux eruptions more visible. However, the `wiggling' of the light curves is insensitive to the change in $R_{\rm Low}$ when $R_{\rm Low}$ is large. The black hole might be in its quiescent state. \label{fig:a0}}
\end{figure*}

\begin{figure}[htb!]
    \centering
    \gridline{
    \fig{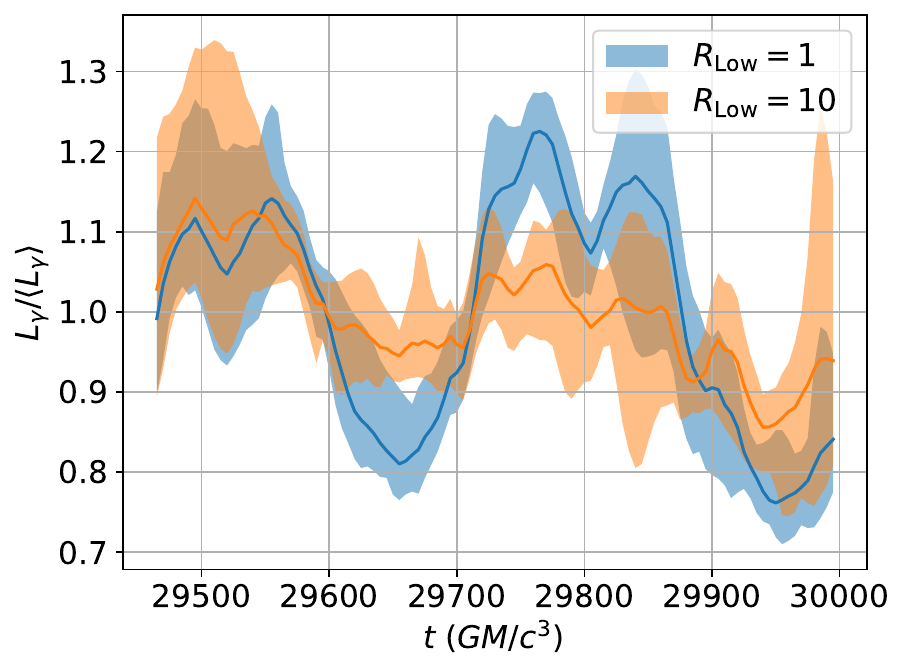}{0.48\textwidth}{(a)}}
    \gridline{
    \fig{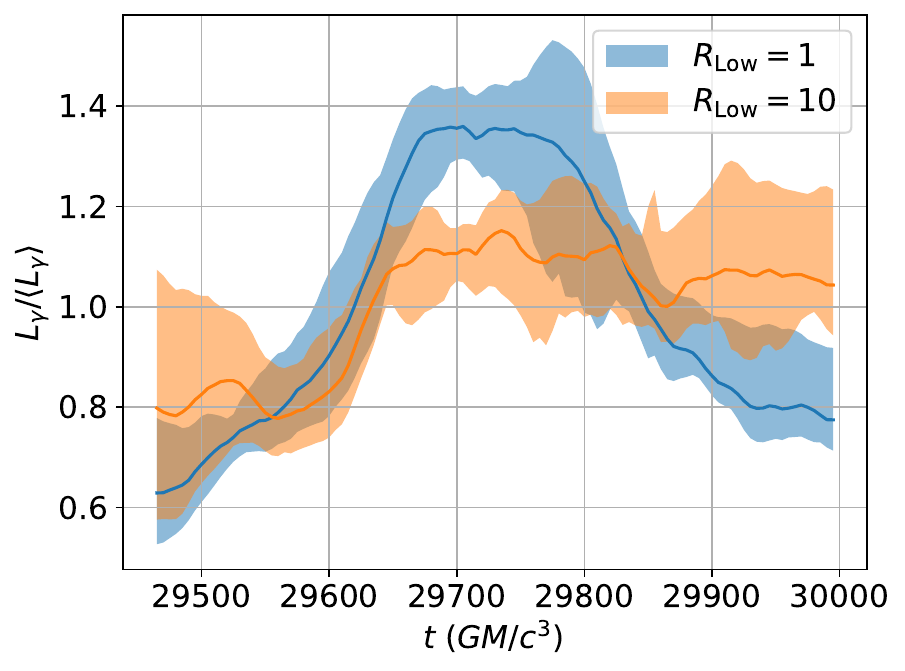}{0.48\textwidth}{(b)}}
    \caption{Transformed light curves for both $R_{\rm Low} = 1, 10$ are shown for models with (a) $a = +0.94$ and (b) $a = +0.5$. The solid line represents the sample mean, while the shaded region indicates the sample range. Increasing $R_{\rm Low}$ leads to an overall `flattening' of the transformed light curves, indicating a reduced variability. \label{fig:varlcs}}
\end{figure}

\begin{figure*}[htb!]
    \centering
    \centering
    \gridline{
    \fig{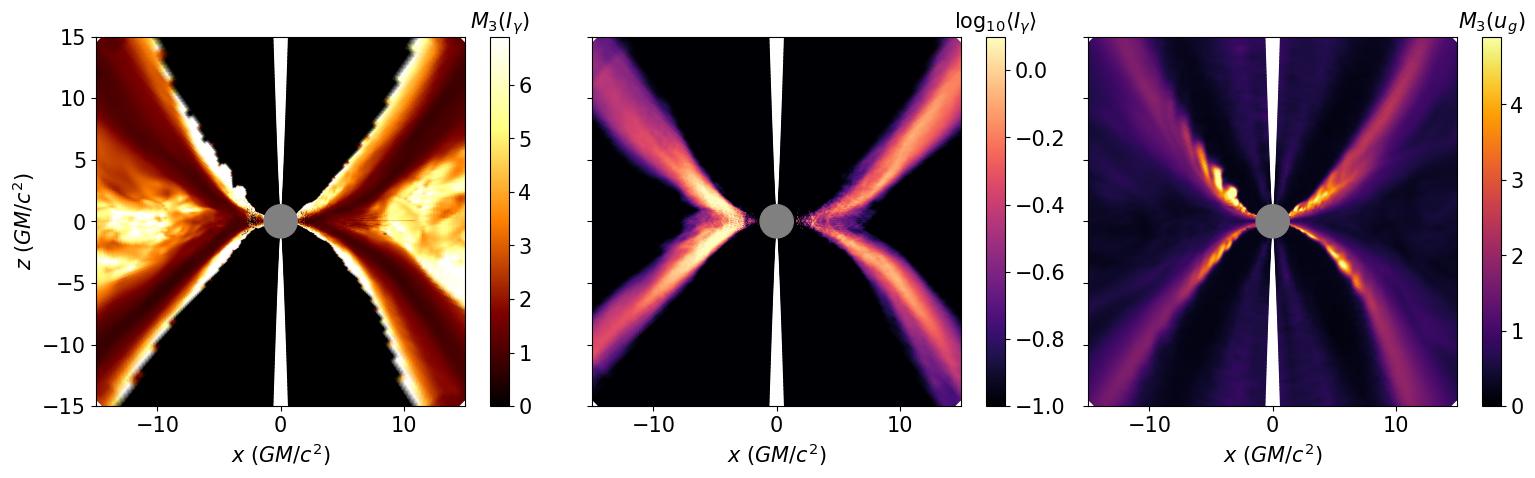}{1.0\textwidth}{(a)}}
    \gridline{
    \fig{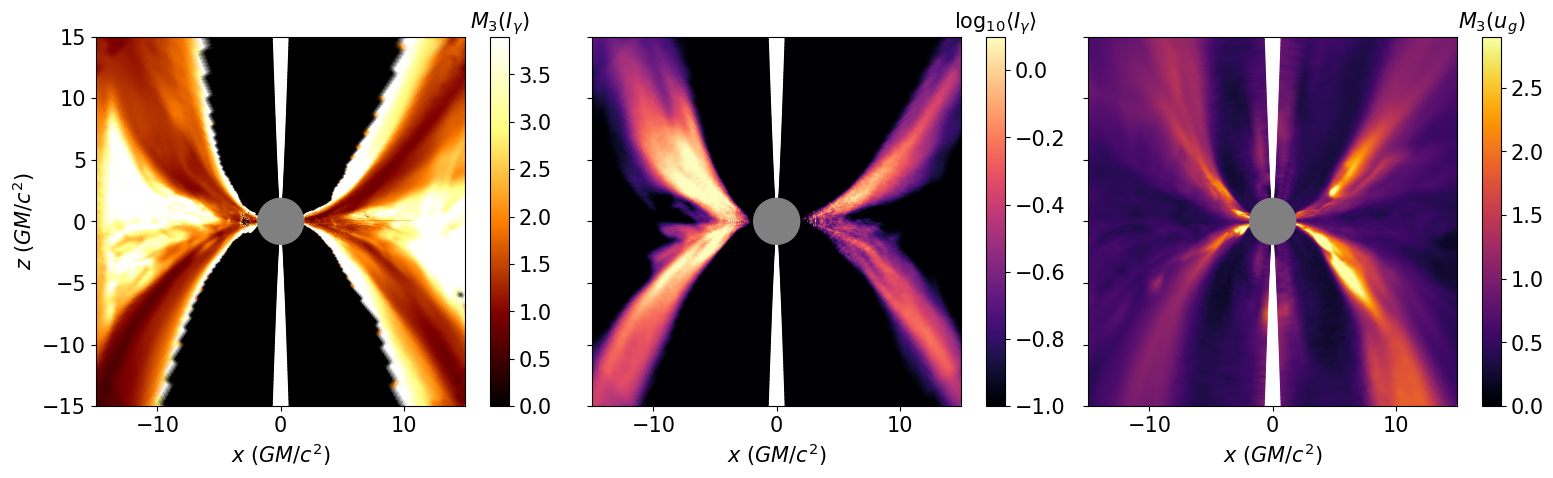}{1.0\textwidth}{(b)}}
    \caption{Computational domain-wise modulation index of the gas internal energy ($M_{3}(u_{g})$), binned emission ($M_{3}(I_{\gamma})$), and the time-averaged binned emission ($\langle I_{\gamma} \rangle$, in the log$_{10}$ scale). Here, $\langle I_{\gamma} \rangle$ is in arbitrary unit. We show models of $a = +0.94$ (a) and $a = +0.5$ (b). The origin of emission is predominantly along the jet funnel wall and is also variable at the same position. The gas internal energy is also most variable along the jet funnel wall. \label{fig:rhovars}}
\end{figure*}

%%%%%%%%%%%%%%%%%%%%%%%%%%%%%%%%%%%%%%%%%%%%%%%%%%%%%%%%%%%%%%%%%%%%%%%%%%%%%%%%%%%%%%%%%%%%%%%%%%%%%%%%%%%%%%%%%%%%%

\subsection{GRRT} \label{sec:grrt}

We raytrace the GRMHD simulation snapshots and produce $230$\,GHz images using the GRRT code \texttt{IPOLE} \citep{moscibrodzka2018ipole}. We solve only the unpolarized radiative transfer equation because the $Q$, $U$, and $V$ parameters are out of scope in this study, but they will be included in a subsequent paper. The unpolarized, covariant radiative transfer equation reads:
\begin{equation} \label{eqn:transfer}
    \frac{d}{d\lambda}\left(\frac{I_{\nu}}{\nu^{3}}\right) = \frac{j_{\nu}}{\nu^{2}} - \alpha_{\nu}\frac{I_{\nu}}{\nu^{2}},
\end{equation}
where the subscript $\nu$ refers to the quantity for a given photon frequency, $I_{\nu}$ is the photon intensity, $j_{\nu}$ is the emissivity, and $\alpha_{\nu}$ is the absorptivity. We assume photons travel along the geodesic, where the geodesic equation is given as:
\begin{align}
     \frac{dx^{\nu}}{d\lambda} &= k^{\nu}, \\
     \frac{dk^{\nu}}{d\lambda} &= -k^{\mu}\Gamma^{\nu}_{\;\;\mu \sigma}k^{\sigma}.
\end{align}
here, $\lambda$ is the affine parameter, and $k^{\nu}$ is the photon four-vector. We use the thermal synchrotron transfer coefficients in \citet{dexter2016public} to solve Equation~\ref{eqn:transfer}. The fixed parameters are as follows \citep{2022ApJ...930L..16E}: The field of view (FOV) is set to be $200$ $\mu$as. The resolution is $400 \times 400$ pixels. The distance from Earth to Sgr~A* is taken as $8,178$ parsecs, while the mass of Sgr~A* is assumed to be $4.154 \times 10^{6}\,M_{\odot}$ \citep{abuter2019geometric}. Note that emission from plasma with $\sigma > 1$ are excluded because of artificial mass injection due to density floors in the simulations.

We perform GRRT parameter surveys on the open science grid \citep{https://doi.org/10.21231/906p-4d78, osg07, osg09} to construct the $230$\,GHz images and $M_{\Delta T}$ with varying $R_{\text{Low}}$. We list the parameters of interest in Table~\ref{tab:params}. They include the spin $a$, the inclination angle of the observer $\theta$, $R_{\rm Low}$, and $R_{\rm High}$. $M_{\rm Unit}$ is a free parameter that scales the gas density of the system and converts from the code unit to the cgs unit. We consider 10 sets of $M_{\rm Unit}$ ranging from $10^{17}$ to $10^{21}$ (in the log scale), and we consider 10 sets of $R_{\rm Low}$ for a given $R_{\rm High}$, ranging from $R_{\rm Low} = 1$ to $R_{\rm Low} = \text{MIN}(60, R_{\rm High})$. We set an upper limit for $R_{\rm Low}$ because an increase in $R_{\rm Low}$ would be associated with an increase in $M_{\rm Unit}$, provided that we fix the $230$\,GHz flux. Thus, we limit $R_{\rm Low}$ so that the assumption of Sgr~A* being optically thin is valid. Finally, we raytrace GRMHD snapshots in the time interval of $\tau = (29,465$ -- $29,995)\,GMc^{-3}$, which covers a duration of $530\,GMc^{-3} \sim 3\,\mathrm{hours}$. This interval represents the last sets of snapshots of the simulations, in which inflow equilibrium should be established beyond the radius where the $230$\,GHz emissions are dominated. We then constrain $M_{\rm Unit}$ and $R_{\rm Low}$ by requiring the time-averaged flux to be $2.4$\,Jy, the observed Sgr~A* $230$\,GHz mean flux \citep{2022ApJ...930L..19W}. Given a $R_{\rm High}$, $\theta$, and $a$, we then have a sequence of models of varying $R_{\rm Low}$ with an associated $M_{\Delta T}$.

%
%%%%%%%%%%%%%%%%%%%%%%%%%%%%%%%%%%%%%%%%%%%%%%%%%%%%%%%%%%%%%%%%%%%%%%%%%%%%%%%%%%%%%%%%%%%%%%%%%%%%%%%%%%%%%%%%%%%%%
%
\begin{deluxetable}{cc}
\caption{List of the parameters of interest, recapped from \citet{2024ApJ...964...17C}. \label{tab:params}}
\tablewidth{0pt}
\tablehead{
\colhead{Parameters} & \colhead{Range}
}
    \startdata
    Black-hole Spin $a$ & ($0.94$, $0.5$, $0$, $-0.5$, $-0.94$) \\
    R$_{\rm High}$ & ($10$, $50$, $90$, $120$, $180$) \\
    Inclination $\theta$ (Deg) & ($0.1$, $45$, $90$, $135$, $179.9$) \\
    $R_{\rm Low}$ & ($1$ -- $\text{MIN}[60, R_{\rm High}]$) \\
    $M_{\rm Unit}$ & ($10^{17}$ -- $10^{21}$)
    \enddata
\end{deluxetable}
%
%%%%%%%%%%%%%%%%%%%%%%%%%%%%%%%%%%%%%%%%%%%%%%%%%%%%%%%%%%%%%%%%%%%%%%%%%%%%%%%%%%%%%%%%%%%%%%%%%%%%%%%%%%
%
\begin{deluxetable}{cccl}[htb!]
\tablecaption{List of black hole parameters for our representative models. \label{tab:models}}
\tablewidth{0pt}
\tablehead{
\colhead{$a$} & \colhead{$\theta$} & \colhead{$R_{\rm High}$} & \colhead{$R_{\rm Low}$} 
}
\startdata
$+0.94$ & $0.1$ & $180$ & ($1$, $14.58$, $37.32$, $60$) \\
$+0.50$ & $45$ & $120$ & ($1$, $23.68$, $41.87$, $60$) \\
$0$ & $135$ & $50$ & ($1$, $33.72$, $41.86$, $50$) \\
$-0.50$ & $0.1$ & $90$ & ($1$, $6.2$, $33.13$, $60$) \\
$-0.50$ & $90$ & $90$ & ($1$, $20.67$, $40.33$, $60$) \\
$-0.94$ & $179.9$ & $180$ & ($1$, $20.67$, $40.33$, $60$) 
\enddata
\end{deluxetable}
%
%%%%%%%%%%%%%%%%%%%%%%%%%%%%%%%%%%%%%%%%%%%%%%%%%%%%%%%%%%%%%%%%%%%%%%%%%%%%%%%%%%%%%%%%%%%%%%%%%%%%%%%%%%%%%%%%%%%%%
%
\subsection{Model Selections} \label{sec:models}

The time interval of interest consists of $107$ GRMHD snapshots. Our GRRT spans $10$ $R_{\rm Low}$ and $10$ $M_{\rm Unit}$ for a given $R_{\rm High}$, and we considered $5$ $R_{\rm High}$ and $5$ $\theta$ for a given $a$. The simulation libraries span $5$ different spins, thus we have approximately $12,500$ $\times$ $107$ = $1,337,500$ GRRT images. It is almost impractical to analyze every single one of the images. Instead, we choose representative models that could help us infer qualitatively the parameter dependence of $M_{\Delta T}$ on $R_{\rm Low}$. We summarize the black hole parameters of our representative models in Table \ref{tab:models} for reference.

The selection procedure is as follows: Given a $R_{\rm High}$, $\theta$, and $a$, we first identify if the sequence of models contains a local minimum in $M_{\Delta T}$. If it does (e.g., upper left and upper center of Figure \ref{fig:paper1recap}), we select 4 representative models along the sequence—one at $R_{\rm Low} = 1$, one at $\text{MIN}[60, R_{\rm High}]$, one at the local minimum, and one in between $\text{MIN}[60, R_{\rm High}]$ and the local minimum. If, however, the sequence of models does not contain a local minimum (e.g., lower center and lower right of Figure \ref{fig:paper1recap}), we will sample 4 models evenly from $R_{\rm Low} = 1$ to $\text{MIN}[60, R_{\rm High}]$.

%
%%%%%%%%%%%%%%%%%%%%%%%%%%%%%%%%%%%%%%%%%%%%%%%%%%%%%%%%%%%%%%%%%%%%%%%%%%%%%%%%%%%%%%%%%%%%%%%%%%%%%%%%%%%%%%%%%%%%%
%
\section{Results} \label{sec:results}
%
%%%%%%%%%%%%%%%%%%%%%%%%%%%%%%%%%%%%%%%%%%%%%%%%%%%%%%%%%%%%%%%%%%%%%%%%%%%%%%%%%%%%%%%%%%%%%%%%%%%%%%%%%%%%%%%%%%%%%
%
\subsection{Positive Spins} \label{sec:positive}

In \citet{2024ApJ...964...17C}, we find that $M_{\Delta T}$ could be reduced by varying $R_{\rm Low}$ for almost all black holes with $a > 0$, and that a local minimum exists between $R_{\rm Low} = 1$ and $\text{MIN}[60, R_{\rm High}]$. To understand this, we show the images and $230$\,GHz light curves of the black hole with $a = +0.94$, $\theta = 0.1$ Deg, and $R_{\rm High} = 180$ in Figure \ref{fig:a+0.94}. We convert images from units of intensity to brightness temperature via:
\begin{equation}
    T_{b} = \frac{I_{\nu}c^{2}}{2k\nu^{2}}
\end{equation}
where $I_{\nu}$ is the intensity of each pixel, $\nu = 230$\,GHz, and $k$ is the Boltzmann constant. The model with $R_{\rm Low} = 1$ has $M_{\Delta T} = 0.147$, while the model with the lowest $M_{\Delta T}$ has $R_{\rm Low} = 14.58$ and $M_{\Delta T} = 0.04$. At a higher $R_{\rm Low}$, $M_{\Delta T}$ increases from the local minimum to a value of $0.180$, which is even larger than that at $R_{\rm Low} = 1$. This is consistent with the `first decrease then increase' trend we found in \citet{2024ApJ...964...17C}. The light curves could be easily separated into different domains in time where the $R_{\rm Low} = 1$ model is more/less variable than models with higher $R_{\rm Low}$. The model with $R_{\rm Low} = 14.58$ has the flattest light curve.

Why does the $R_{\rm Low} = 14.58$ model have the lowest variability? And where does the large variability of the higher $R_{\rm Low}$ models originate? We select GRRT snapshots at $t = 29,840$\,$GM/c^{3}$ and show them on the first row of Figure \ref{fig:a+0.94}. They are marked as squares on the light curves and correspond to the moment where the $R_{\rm Low} = 1$ model is more variable than the rest. The major contribution to the $230$\,GHz flux is from the photon ring, consistent with our findings in \citet{2024ApJ...964...17C}. As $R_{\rm Low}$ increases, the optical depth of the fluid also increases and helps obscure the photon rings, reducing its contribution to the variability. We also note that the higher $R_{\rm Low}$ models at this particular moment are dimmer than the averaged $230$\,GHz flux. We illustrate the increase in optical depth with increasing $R_{\rm Low}$ by showing the time-averaged optical depth of the black hole in Figure \ref{fig:odepth}. The frequency-dependent optical depth along each ray that follows the geodesic is given as an integral:
\begin{equation}
  \tau_{\nu} = \int_{\lambda_0}^{\lambda 1}\nu\alpha_{\nu}d\lambda
\end{equation}
where $\lambda_0$ ($\lambda_1$) is the point of emission (recipient). Note that the absorption coefficient $\alpha_{\nu}$ is evaluated in the fluid rest frame. We compute optical depth images for each snapshot and then take the time average. Here, the scale $1 - 2$ is mapped, while that from $0 - 1$ is physical. The $R_{\rm Low} = 1$ model has an optically thin photon ring. The photon ring for the $R_{\rm Low} = 14.58$ model is marginally optically thin, while those with larger $R_{\rm Low}$ are optically thick. These combined results suggest that the reduced variability is an optical depth effect -- the gravitationally lensed photons, even though traveling through the same physical length, sample a larger photon path length due to increased fluid density.

We show in the second row of Figure \ref{fig:a+0.94} the images at $t = 29,980$\,$GM/c^{3}$, where the higher $R_{\rm Low}$ models are more variable than the $R_{\rm Low} = 1$ and $14.58$ models. These are marked as triangles on the light curves. In the image plane, we find that the higher $R_{\rm Low}$ models show bright transients near the black hole, which become more visible as $R_{\rm Low}$ increases. The origins of this transient can be understood through analyzing GRMHD snapshots. We plot the plasma entropy $\mathcal{S} = \text{log}\{ P/[\rho^{\gamma}(\gamma - 1)] \}$ contours at this particular moment in Figure \ref{fig:eruption}. We find a patch of plasma close to the black hole that has an unusually large $\mathcal{S}$ compared to its surroundings, and it has a shape consistent with the bright transients shown in Figure \ref{fig:eruption}. The $x-z$ slice of the snapshot shows a tube-like structure for this patch of plasma. This is probably a MAD flux eruption caused by reconnections \citep{dexter2020sgr, scepi2022sgr, scepi2024magnetic}.

Black holes with $a = +0.5$ exhibit similar trends in the $R_{\rm Low}$ vs. $M_{\Delta T}$ curves as those with $a = +0.94$. To understand this, we show the $230$\,GHz light curves and images for black holes with $a = +0.5$, $\theta = 45$ Deg, and $R_{\rm High} = 120$ in Figure \ref{fig:a+0.5}. Like Figure \ref{fig:a+0.94}, we can divide the light curves into domains where the $R_{\rm Low} = 1$ models are more or less variable than the remaining ones. The model with $R_{\rm Low} = 23.68$ is the least variable with $M_{\Delta T} = 0.103$, and its light curve is the flattest.

We show the images in the first row at $t = 29,780$\,$GM/c^{3}$ (marked as triangles on the light curves) where the $R_{\rm Low} = 1$ model is more variable than the others. We find, again, that the major contribution to the brightness for the $R_{\rm Low} = 1$ model is from the photon ring and that optically thick fluid blocks the photon ring from illuminating for models with a higher $R_{\rm Low}$.

At the beginning of this particular time chunk, the higher $R_{\rm Low}$ models are much more variable. We show in the second row of Figure \ref{fig:a+0.5} the images at this instant. We find a prominent, tube-like feature in the images, which becomes more visible as $R_{\rm Low}$ increases further. This is probably a large-scale MHD flux tube created by the MAD flux eruptions --- after the black hole saturates in horizon-penetrating magnetic flux, part of the magnetic flux is dissipated via a magnetic reconnection that generates a strong vertical magnetic field, which then pushes back the accretion flow and creates the flux tube \citep{vos2023magnetic}. We illustrate the image of the flux tube viewed at different $\theta$ in Figure \ref{fig:fluxtube}. Note that $R_{\rm Low} \neq 1$ means the image is slightly optically thicker than the $R_{\rm Low} = 1$ model. In Figure \ref{fig:fluxtube-vecfield}, we show the density contour with magnetic field lines at the same instant. A patch of tube-like, low-density plasma threaded with strong vertical magnetic fields is visible, thus confirming our claim that the tube-like structure is a MAD eruption flux tube.

In short, for black holes with $a > 0$, the major contribution to the variability of the $R_{\rm Low} = 1$ model is from the photon ring. Increasing $R_{\rm Low}$ increases the optical depth of the fluid, which helps block the photon ring from illuminating. However, this comes with the cost of making MAD flux eruptions more visible, thus increasing $M_{\Delta T}$. The sweet spot occurs when the contributions to $M_{\Delta T}$ from the photon rings and the flux eruptions are \textit{combined} at the minimal.

%%%%%%%%%%%%%%%%%%%%%%%%%%%%%%%%%%%%%%%%%%%%%%%%%%%%%%%%%%%%%%%%%%%%%%%%%%%%%%%%%%%%%%%%%%%%%%%%%%%%%%%%%%%%%%%%%%%%%

\subsection{Negative Spins} \label{sec:negative}

In \citet{2024ApJ...964...17C}, we observed an increasing trend in $M_{\Delta T}$ when we attempted to increase $R_{\rm Low}$ for almost all black holes with $a < 0$. To understand this, we show the $230$\,GHz light curves and images in Figure \ref{fig:a-0.5}. Here, the black hole parameters are $a = -0.5$, $\theta = 0.1$ Deg, and $R_{\rm High} = 90$. We show only the moment where the higher $R_{\rm Low}$ models are more variable. We find that the major contributions to the variability for the $R_{\rm Low} = 1$ models are from the photon ring, consistent with our findings in \citet{2024ApJ...964...17C}. Increasing $R_{\rm Low}$ also makes the fluid optically thicker, as in the $a > 0$ cases, but the fluid contains voids that cannot cover the entire photon ring. We also show the black holes with the same spin but viewed at $\theta = 90$ Deg in Figure \ref{fig:a-0.5-angle}. We see the same image morphology where optically thick gas cannot cover the photon ring. Thus, this phenomenon should be generic across varying $\theta$. The origin of these voids could be easily traced through GRMHD snapshots. We show the $x-y$ ($x-z$) density contours in the upper (lower) panel of Figure \ref{fig:void}. We observe a flux eruption events as in the $a > 0$ case, and we also find patches of low-density plasma surrounding the black hole.

As in the $a > 0$ cases, MAD flux eruptions are more visible as one increases $R_{\rm Low}$, which can be seen as bright strips in the increasing $R_{\rm Low}$ models in Figure \ref{fig:a-0.5}. We verify this by showing the entropy $\mathcal{S}$ contour in Figure \ref{fig:eruption-a-0.5}. We find a strip of unusually high $\mathcal{S}$ plasma floating away from the black hole, which has a shape consistent with the bright strips shown in Figure \ref{fig:a-0.5}. We note that the flux eruptions span a larger spatial scale, making them visible even when $R_{\rm Low} = 1$. This differs from the $a > 0$ cases, where the flux eruptions become visible only when $R_{\rm Low} > 1$ and their spatial scale is much smaller.

There is an exception to the above discussion: the model with $a = -0.94$ and $\theta = 179.9$ Deg. We find a decreasing trend in the $R_{\rm Low}$ vs. $M_{\Delta T}$ curves regardless of the $R_{\rm High}$ assumed. To understand this, we show the images and $230$\,GHz light curves of black holes for this set of parameters but with $R_{\rm High} = 180$ in Figure \ref{fig:a-0.94}. The photon rings are still visible for models with $R_{\rm Low} > 1$. The flux eruptions are visible even at $R_{\rm Low} = 1$. The combined variability due to flux eruptions and photon rings makes the $R_{\rm Low} = 1$ model stand out compared to those with $R_{\rm Low} > 1$. However, since the increasing trend is quite generic for all of the $a = -0.5$ models and for the $a = -0.94$ models viewed from another angle, we believe that this is a particular case that would not affect our qualitative understanding of the parameter dependence of $M_{\Delta T}$ on $R_{\rm Low}$ for black holes with $a < 0$.

%%%%%%%%%%%%%%%%%%%%%%%%%%%%%%%%%%%%%%%%%%%%%%%%%%%%%%%%%%%%%%%%%%%%%%%%%%%%%%%%%%%%%%%%%%%%%%%%%%%%%%%%%%%%%%%%%%%%%

\subsection{Zero Spin} \label{sec:zero}

We find that $M_{\Delta T}$ is not sensitive to the variation of $R_{\rm Low}$ for black holes with $a = 0$, irrespective of the $R_{\rm High}$ assumed. Here, we select the black holes with $a = 0$, $\theta = 135$ Deg, and $R_{\rm High} = 50$ and show the resulting $230$\,GHz light curves and images in Figure \ref{fig:a0}. The optical depth argument still applies, in which the photon rings are blocked by optically thick gas. However, the $230$\,GHz light curves for models with larger $R_{\rm Low}$ are similar, even though they differ significantly from that of $R_{\rm Low} = 1$. The light curves thus reveal why $M_{\Delta T}$ is insensitive to $R_{\rm Low}$. The reason for this behavior might be that the interval of snapshots we selected is when the black holes are at their quiescent state, or the black holes with $a = 0$ are intrinsically quiet. Nonetheless, given that the current EHT constraint on Sgr~A* prefers a black hole with a non-zero spin, the $a = 0$ case is seemingly out of interest and would not affect our physical interpretation of the variability of modeling Sgr~A*. However, the physics of such insensitivity would be interesting, and we will leave this work for a subsequent paper.

%%%%%%%%%%%%%%%%%%%%%%%%%%%%%%%%%%%%%%%%%%%%%%%%%%%%%%%%%%%%%%%%%%%%%%%%%%%%%%%%%%%%%%%%%%%%%%%%%%%%%%%%%%%%%%%%%%%%%

\section{Discussion} \label{sec:discuss}

%%%%%%%%%%%%%%%%%%%%%%%%%%%%%%%%%%%%%%%%%%%%%%%%%%%%%%%%%%%%%%%%%%%%%%%%%%%%%%%%%%%%%%%%%%%%%%%%%%%%%%%%%%%%%%%%%%%%%

%\begin{figure}[htb!]
%    \centering
%    \includegraphics[width=1.0\linewidth]{corr-overall.pdf}
%    \caption{The averaged correlation coefficient was computed between the %total gas internal energy $U_{\rm tot}(r)$, integrated within a volume of %radius $r$, and all the $R_{\rm Low} = 1$ light curves. The maximum correlation %coefficient exceeds $0.6$ for all spins, indicating a moderately positive %correlation. \label{fig:corr-overall}}
%\end{figure}
%
%\begin{figure*}[htb!]
%    \centering
%    \includegraphics[width=1.0\linewidth]{corr-lc-all.pdf}
%    \caption{The rescaled time series of $U_{\rm tot}$ and $L_{\gamma}$ that is %computed at $R_{\rm Low} = 1$. Here, the $U_{\rm tot}$ time series is shown as %a black line, corresponding to the point of maximum correlation coefficient %shown in Figure \ref{fig:corr-overall}. $L_{\gamma}$ is shown as blue dotted %lines. The `wiggling' of $U_{\rm tot}$ moderately follows that of $L_{\gamma}$. %\label{fig:corr-lc}}
%\end{figure*}

%%%%%%%%%%%%%%%%%%%%%%%%%%%%%%%%%%%%%%%%%%%%%%%%%%%%%%%%%%%%%%%%%%%%%%%%%%%%%%%%%%%%%%%%%%%%%%%%%%%%%%%%%%%%%%%%%%%%%

In \citet{2024ApJ...964...17C}, we conjectured that the high $M_{\Delta T}$ of numerical models is caused by the high sensitivity of the electron temperature prescription (Equation \ref{eqn:fraction}) to a slightly changing $\beta$ within $\beta = 10^{-1} - 10^{1}$. Thus, if one can close the gap between $R_{\rm Low}$ and $R_{\rm High}$, such sensitivity could be reduced and potentially help reduce $M_{\Delta T}$. However, in this study, even though we find that increasing $R_{\rm Low}$ could reduce $M_{\Delta T}$ for black holes with $a > 0$, it is merely an optical depth effect --- gas becomes optically thick and blocks the photon rings from illuminating. Nonetheless, we do not rule out the possibility that the shape of Equation \ref{eqn:fraction} plays an important role in causing the high $M_{\Delta T}$ when the photon rings are optically thin. Without changing $R_{\rm Low}$, one could easily tune Equation \ref{eqn:fraction} so that the transition between $R_{\rm Low}$ and $R_{\rm High}$ has a flatter slope, and we leave this study for the future.

We have shown that the reduction of $M_{\Delta T}$ due to the increase in $R_{\rm Low}$ is more prominent for models with $a > 0$, with a few particular cases. It is interesting to examine if such a reduction is systematic across samples, provided that one does not increase $R_{\rm Low}$ excessively. To achieve this, we extract light curves $L_{\gamma}$ for $R_{\rm Low} = 1$ and $10$ across all $R_{\rm High}$ and $\theta$, focusing on $a = +0.94$ and $0.5$. Specifically, we perform the transformation $L_{\gamma} \rightarrow L_{\gamma}/\langle L_{\gamma} \rangle$ so that the `wiggling' of the light curve with respect to $1$ represents its time-variability. We present the results in Figure \ref{fig:varlcs}. We observed an overall flattening of the sample- mean and range of the transformed light curves, indicating that the effect of reducing $M_{\Delta T}$ when one increases $R_{\rm Low}$ is not a coincidence, at least for models with $a > 0$.

We also find that increasing $R_{\rm Low}$ would come with the cost of making MAD flux eruptions more visible and contribute to a higher $M_{\Delta T}$, especially for black holes with $a < 0$. However, given that MAD models with $a > 0$ are more favored \citep{2022ApJ...930L..16E}, we believe that varying $R_{\rm Low}$ would at least be an intermediate solution to the variability crisis. In the longer term, one should address the origin of the variabilities. We guess it is related to the time-variability of the gas internal energy. To test our idea, we extract the binned location of the point of origin for all photons that make up an image $I_{\gamma}$ (see Figure 4 of \citet{akiyama2019first}) and the gas internal energy $u_{g} = P/(\gamma - 1)$. Specifically, we compute the modulation index of these two quantities, defined as $\sqrt{\text{Var}(Q)}/\langle Q \rangle$ for the time interval $\tau = (29,465$ -- $29,995)\,GMc^{-3}$ in a computational domain-wise fashion. We also compute the time-average of $I_{\gamma}$. The domain-wise modulation index and time-average of $I_{\gamma}$ are then further averaged across the samples. We present the results in Figure \ref{fig:rhovars}, focusing on models with $a > 0$. We find that the emissions are predominantly along the jet funnel wall, which is also the location where the emission is variable. Additionally, we observe that the internal energy is most variable along the jet funnel wall, and in some instances, the domain-wise modulation index of $u_{g}$ has a similar shape to that of $I_{\gamma}$. These observations suggest that the time-variability of the light curves may correlate with the time-variability of the gas internal energy.

Suppose the time variability of the gas internal energy is the major contribution to the time variability of the light curves. In that case, its source is likely plasma heating, which directly impacts the plasma's internal energy. The source of heating could be grid-scale dissipation. To address the variability crisis with more accurate modeling of plasma heating, one should carefully tune the simulation parameters to control the frequency of, or the energy released by, these heating events, if one aims to reduce $M_{\Delta T}$. The solution should contain two key components: (i) simulations that better control the amount of dissipative heating, and (ii) a better electron temperature prescription that regulates the response of electron temperature to the heating. To better control the amount/frequency of dissipative heating, one could consider using less diffusive numerical schemes or more first-principle modeling of dissipation, such as viscous GRMHD or resistive GRMHD. Additionally, one should look for an electron heating function that does not respond rapidly to changes in $u_{g}$.

Another reason for the high time variability of numerical models could be due to the $\sigma$ cutoff. We show that the source of the light curve time-variability is mostly along the jet funnel wall, while all the emission inside the funnel wall is set to zero by the default $\sigma = 1$ cut. If the $\sigma = 1$ boundary is highly variable, it is evident that the emission is also highly variable due to the steep step function of emission at $\sigma = 1$. If this is the major cause of the variability, one could consider smoothing the emission transition at $\sigma = 1$ using an analytic function such as the logistic function. Lastly we should remark that the $M_{3}$ maps we shown in Figure \ref{fig:rhovars} might subject to spatial-variability, and thus might not reflect the time-variability of the lightcurves, which are summation of the all emission and thus should be less sensitive to the spatially varying emission point (i.e., same emission strength but located at different position at different time). Lastly, we note that in GRMHD models, the evolution of the gas internal energy near the $\sigma = 1$ interface remains uncertain due to significant density gradients, which could also contribute to the increased variability.
 
%%%%%%%%%%%%%%%%%%%%%%%%%%%%%%%%%%%%%%%%%%%%%%%%%%%%%%%%%%%%%%%%%%%%%%%%%%%%%%%%%%%%%%%%%%%%%%%%%%%%%%%%%%%%%%%%%%%%%

\section{Conclusion} \label{sec:conclu}

The Event Horizon Telescope Collaboration has successfully observed the galactic center black hole Sgr~A*. Model comparison with observational data provided valuable constraints on the black hole properties, such as spin, observer inclination, plasma temperature, accretion mode, etc. However, current GRMHD models are inconsistent with the 3-hour, $230$\,GHz flux variability ($M_{\Delta T}$) constraint \citep{2022ApJ...930L..16E}, thus creating the variability crisis for modeling Sgr~A*. In \citet{2024ApJ...964...17C}, we addressed this problem by varying one of the fixed parameters in the ion-electron temperature ratio prescription functions (Equation \ref{eqn:fraction}) --- $R_{\rm Low}$, representing the ion-electron temperature ratio in the strongly magnetized regime. We found that increasing $R_{\rm Low}$ beyond 1 helps reduce $M_{\Delta T}$ for almost all black holes with $a > 0$, but this is not the case for almost all black holes with $a < 0$. The variation of $M_{\Delta T}$ is insensitive to changes in $R_{\rm Low}$ for black holes with $a = 0$. In this study, we aim to address the parameter dependence of $M_{\Delta T}$ on increasing $R_{\rm Low}$ through analyzing suites of GRMHD simulation and GRRT snapshots. We select representative models for each black hole spin and examine the origin of variabilities as seen in the $230$\,GHz flux light curves. Our major findings are:
\begin{enumerate}

    \item For black holes with $a > 0$, we find that the variability for the $R_{\rm Low} = 1$ models originates from the photon rings, which is consistent with our findings in \citet{2024ApJ...964...17C}. As $R_{\rm Low}$ increases, the fluid becomes more optically thick, blocking and preventing the photon ring from illuminating, thus reducing $M_{\Delta T}$. Therefore, reducing $M_{\Delta T}$ due to increasing $R_{\rm Low}$ is merely an optical depth effect. As $R_{\rm Low}$ is further increased, MAD flux eruptions become more visible and help offset the reduced variability of the obstructed photon rings. An optimal solution for the lowest $M_{\Delta T}$ occurs when the variability due to both photon rings and MAD flux eruption events are combined to a minimum. 
    
    \item We observe the same phenomenon for black holes with spin $a < 0$. However, the optically thick fluid contains voids that cannot cover the photon rings. Additionally, flux eruptions have a larger spatial scale and are thus more visible. Hence, the increase in the optical depth accompanied by the rise of $R_{\rm Low}$ does not substantially reduce variabilities originating from the photon rings and, on the other hand, enhances variabilities due to MAD flux eruptions. We note that models with $a = -0.94$ and $\theta = 179.9$ Deg show a decrease in $M_{\Delta T}$ when $R_{\rm Low}$ is increased. We argue that this is a particular case. Since the increasing trend in $M_{\Delta T}$ is quite generic across black holes with $a < 0$, this particular case does not affect our qualitative understanding of the parameter dependence of $M_{\Delta T}$ on $R_{\rm Low}$. 
    
    \item The optical depth effect still applies to black holes with spin $a = 0$, but their light curves are similar across models with $R_{\rm Low} \neq 1$, which reveals why $M_{\Delta T}$ is not very sensitive to the variations in $R_{\rm Low}$. We suspect this is because either we selected an interval of snapshots where the black hole is in its quiescent state, or the black hole with $a = 0$ is intrinsically quiet. Parameter studies with a longer time interval of snapshots are required to understand such insensitivity. However, given that current constraints on the spin of Sgr~A* prefer $a \neq 0$, we believe that this would not affect our qualitative understanding of the physics of changing $M_{\Delta T}$ by increasing $R_{\rm Low}$, at least within the parameter space of interest. 
    
    \item We also briefly discuss the physical origin of the variability of the lightcurves at $R_{\rm Low} = 1$, the optically thin limit. We show that the origin of emission is predominantly along the jet funnel wall and is also variable at the same position, which is where the gas internal energy is also the most variable. Thus, we suspect that the time variability of gas internal energy is related to that of the light curves. Grid-scale dissipative heating could be the source of the variability of $u_{g}$. We also note that the abrupt $\sigma = 1$ cutoff in emission, along with the uncertainty in the evolution of $u_{g}$ within GRMHD models at the $\sigma = 1$ interface, could serve as additional sources of variability.
    
\end{enumerate}

Last but not least, we stress that increasing $R_{\rm Low}$ is an intermediate solution that allows us to understand the origin of variabilities through the image domain only. In the longer term, one should address why the $R_{\rm Low} = 1$ models are variable. Finally, the current study only considered the effects of $R_{\rm Low}$ as manifested in the intensity domain. Since the reduction in $M_{\Delta T}$ is an optical depth effect, we expect it to be seen through polarimetric signatures. For instance, increasing the optical depth would promote Faraday rotations, increasing the circular and reducing the linear polarization. Comparing resolved or unresolved polarimetric signatures with observations could help test the $R_{\rm Low} \neq 1$ model. Also, a higher optical depth would make the spectral index more positive. Computing either the unresolved spectral index or resolved spectral index map and comparing them with observation could help us constrain the parameter space of $R_{\rm Low}$. We plan to perform these extended studies in our third series of papers.

%(i.e., identify the source of abnormal heating) and control the amount of energy released, as well as the frequency of the heating. One should also place better constraints on the ion-electron temperature prescriptions and/or electron heating functions to control the response of electron temperature to these heating.

%%%%%%%%%%%%%%%%%%%%%%%%%%%%%%%%%%%%%%%%%%%%%%%%%%%%%%%%%%%%%%%%%%%%%%%%%%%%%%%%%%%%%%%%%%%%%%%%%%%%%%%%%%%%%%%%%%%%%

\begin{acknowledgments}
We thank Charles Gammie, Benjamin Prather, and George Wong for providing us the EHT v3 simulation libraries to carry out the raytracing calculations, as well as giving us valuable comments that improves the manuscript. We acknowledge support from the ACCESS allocation AST170024. H.-s.~C. acknowledges support from the Croucher Scholarship for Doctoral Studies by the Croucher Foundation. C.-k.~C. acknowledge NSF grant 2034306 support.
\end{acknowledgments}

%%%%%%%%%%%%%%%%%%%%%%%%%%%%%%%%%%%%%%%%%%%%%%%%%%%%%%%%%%%%%%%%%%%%%%%%%%%%%%%%%%%%%%%%%%%%%%%%%%%%%%%%%%%%%%%%%%%%%

\bibliography{main}{}
\bibliographystyle{aasjournal}

%%%%%%%%%%%%%%%%%%%%%%%%%%%%%%%%%%%%%%%%%%%%%%%%%%%%%%%%%%%%%%%%%%%%%%%%%%%%%%%%%%%%%%%%%%%%%%%%%%%%%%%%%%%%%%%%%%%%%

\end{document}